\documentstyle[aps,prl,epsfig,special]{revtex}

\begin{document}
\draft
\title{Transmission Phase  Shift
of a Quantum Dot with Kondo Correlations}
\author{Ulrich Gerland$^1$, Jan von Delft$^1$, T. A. Costi$^{2,\ast}$ and 
Yuval Oreg$^3$}
\address{$^1$Institut f\"ur Theoretische Festk\"orperphysik, 
Universit\"at Karlsruhe, 76128 Karlsruhe, Germany}
\address{$^2$Theoretische Physik III, Universit\"at Augsburg, 
  86135 Augsburg, Germany}
\address{$^3$Lyman Laboratory of Physics, Harvard University, 
         Cambridge MA 02138, USA}
\date{Submitted November 24, 1999; published in Phys.\ Rev.\ Lett.\
  {\bf 84}, 3710 (2000)}
\maketitle
\begin{abstract}
  We study the effects of Kondo correlations on the transmission phase
  shift of a quantum dot in an Aharonov-Bohm ring. We predict in
  detail how the development of a Kondo resonance should affect the
  dependence of the phase shift on transport voltage, gate voltage and
  temperature.  This system should allow the first direct observation
  of the well-known scattering phase shift of $\pi/2$ expected (but
  not directly measurable in bulk systems) at zero temperature for an
  electron scattering off a spin-$1 \over 2$ impurity that is screened
  into a singlet.
\end{abstract}
\pacs{PACS numbers:
75.20.Hr,  
72.15.Qm,   
73.23.Hk,  
73.20.Dx,  
}
\narrowtext

The Kondo effect for metallic electrons interacting with localized
spins has been studied for more than three decades \cite{hewson-book},
yet one of its most fundamental properties has so far eluded direct
experimental verification: at temperatures sufficiently low that a
spin-${1 \over 2}$ impurity is screened into a singlet, a conduction
electron scattering off the latter is predicted\cite{friedel,langreth}
to suffer a resonance phase shift of $\pi/2$ \cite{nozieres}.
 A direct observation of this phase
shift, not possible in bulk systems, has now become feasible using
quantum dots, due to two recent experimental breakthroughs: Kondo-type
correlations were observed in dots strongly coupled to
leads\cite{goldhaber98a,cronenwett98a,goldhaber98b,schmid98a,simmel98},
and it was demonstrated that 
the transmission phase shift of a dot can be  measured by
Aharonov-Bohm (AB) interferometry\cite{yacoby95,schuster}.  

In \cite{goldhaber98a,cronenwett98a,goldhaber98b,schmid98a,simmel98},
a semi-conductor quantum dot was coupled via tunnel junctions to leads
and capacitively to a gate.  Tuning the gate voltage $V_g$, which
linearly shifts the dot's eigenenergies relative to the chemical
potential of the leads, produced a Coulomb-blockade peak in the dot's
conductance each time its electron number $N$ changed by one.  In
valleys between two peaks in which $N$ is odd and at sufficiently low
temperatures $T$, the conductance showed anomalous features
\cite{goldhaber98a,cronenwett98a,goldhaber98b,schmid98a,simmel98} in
accord with earlier predictions \cite{glazman88a,meir93a,koenig96a}.
These are due to ``Kondo correla\-\mbox{tions}'', which arise when the
dot's topmost (spin-degenerate) occupied energy level, henceforth
called the $d$-level, carries on average a {\em single}\/ electron
that can mimic a magnetic spin-${1 \over 2}$ impurity in a metal,
leading to the well-known Kondo effect\cite{hewson-book}.  Quantum
dots can thus be used as ``tunable Kondo impurities''.  Embedding such
a dot in one arm of an AB interferometer and measuring the {\em
  phase shift\/} of the transmission amplitude through the dot
\cite{yacoby95,schuster} would thus amount to measuring the scattering
phase shift off an Kondo impurity \cite{heiblum}.  
In this Letter we
predict in detail, within the framework of the Anderson model,
how Kondo correlations influence the dot's transmission
phase shift, and explain how the Kondo phase shift of $\pi/2$ should
manifest itself.

\begin{figure}[h]
\begin{center}
    \epsfig{figure=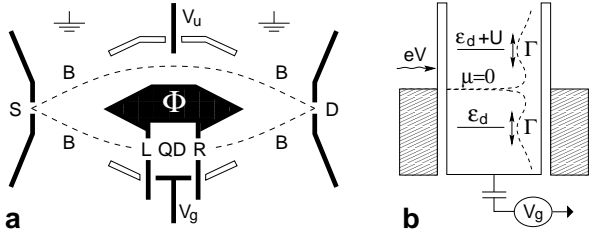,width=0.90\linewidth} \vspace{4mm}
    \caption{
      (a) An AB interferometer: The source S, base region B and drain
      D have chemical potentials $\mu_S = eV$, $\mu = 0$ and $ \mu_D =
      0$, respectively.  In the base region four reflectors (shown in
      white) and a central barrier (black) define an upper and lower arm
      forming a ``ring'' (dotted lines) threaded by an
      applied magnetic flux $\Phi$.  The lower arm contains a quantum
      dot (QD), coupled to the base region via tunable left and right
      tunnel barriers (L,R).  The gate voltages $V_{g{}}$
      or $V_u$ can be used to sweep the dot's energy levels relative
      to $\mu$, or to change the transmission amplitude of the upper
      arm, respectively. (b) Energy diagram of a QD
      whose density of states (dashed line) has a Kondo resonance
      with width of order $T_K$ at energy $E = \mu = 0$, 
      and two broad single-particle resonances,
      with widths $\Gamma$, at $\varepsilon_d < 0$ and $\varepsilon_d
      + U > 0$.  Electrons incident from the source have 
      average energy $eV$ relative to $\mu$.  
\vspace*{-4mm} }
    \label{dot} \label{interferometer}
\end{center}
\end{figure}

{\em AB-interferometry.}---
Fig.~\ref{interferometer}(a) depicts an
AB-in\-ter\-fe\-ro\-meter \cite{schuster}.  
A spin-$\sigma$ electron injected
from the source can reach the drain through both the upper or lower
arm, with transmission amplitudes $t_{u \sigma}$ or $t_{l \sigma}$.
Their phase difference has the form $2 \pi \Phi e/h + \delta \phi $,
where $\Phi$ is the magnetic flux enclosed by the ``ring'' formed by
the arms.  In\-ter\-fe\-rence between $t_{u \sigma}$ and $t_{l
  \sigma}$ causes the
differential conductance $dI/dV$ measured at the drain
to exhibit AB oscillations as function of $\Phi$, which are of
the form\cite{schuster} 
\begin{equation}
  \label{eq:AB-oscillations}
  G_{AB} \propto {e^2 \over h} \sum_\sigma
  | t_{u \sigma}| |t_{l \sigma} | 
  \cos (2 \pi \Phi e/h + \delta\phi ) . 
\end{equation}
The lower arm contains a quantum dot, hence $t_{l \sigma}$ is
proportional to the transmission amplitude $t_{d \sigma}$ through the
dot.  By recording how the amplitude and phase of the AB oscillations
change with gate voltage $V_{g{}}$, source-drain voltage $V$ or
temperature $T$, one can thus measure the dependence on these
parameters of $|t_{d \sigma}|$ and the ``transmission phase shift''
$\phi_{d \sigma} = {\rm arg} (t_{d\sigma}) = \delta \phi + {\rm
  const.}$

To derive an explicit expression for $t_{l \sigma}$, we 
calculated\cite{vD} $G_{AB}$ using the general theory
for AB interferometers of Ref.~\onlinecite{bruder96a}, which assumes
(i) that transport through the ring is fully coherent \cite{coulomb}.
We  further assumed that (ii)~the dot level spacing $\Delta$ is
so large that only the $d$-level influences transport through the
dot\cite{goldhaber98a,cronenwett98a,goldhaber98b,schmid98a,simmel98};
(iii)~the slits of source and drain are so small that only one conducting mode
carries current between them\cite{schuster}; (iv)~multiple traversals of the
ring can be neglected due to the open nature of the base region
\cite{schuster}; and (v)~the source and
drain do not drive the dot out of equilibrium\cite{yacoby-priv}.

The result for $G_{AB}$ is of the form
(\ref{eq:AB-oscillations}), with $t_{l \sigma} = ({\cal N}_0
t_{0\sigma} | t_L t_R| /\Gamma ) t_{d \sigma}$, where ${\cal N}_0$ is
the density of states per mode in the base region (assumed constant),
$t_{0 \sigma}$ is a geometrical factor of order unity depending on the
amplitudes to reach the dot from the source or drain, $t_L$ ($t_R$) is
the amplitude per mode for tunneling between dot and base region
through the left (right) tunnel barrier, and $\Gamma$ is the width
acquired by the $d$-level due to this coupling. All $V_{g{}}$, $V$ and
$T$-dependencies reside in the remaining factor (a $V \neq
0$ generalization of Ref.~\onlinecite{oreg97a}), 
\begin{equation}
\label{transmission}
  t_{d \sigma} (V_{g{}},V,T)  = \Gamma
  \int d E\,\frac{\partial 
f (E - e V) }{\partial E}\,{\cal G}_{d\sigma} (E) \;  ,
\end{equation}
\noindent the ``thermally averaged transmission
amplitude'' through the dot for electrons incident with mean energy
$eV$.  Here $f(E)$ is the Fermi function and ${\cal G}_{d\sigma} (E)$
the retarded Green's function for a spin-$\sigma$ electron on the
$d$-level. ${\cal G}_{d\sigma}$ depends on $V_{g{}}$ (via
$\varepsilon_d$) and on $T$ (due to Kondo correlations), but not
on $V$ (by assumption (v), the  dot is in equilibrium with the
base region).  Since AB oscillations can be produced already with a
magnetic field $H$ much too weak to lift the spin-degeneracy of the
$d$-level, we for the most part shall take ${\cal G}_{d\sigma}$ to be
$H$-independent, so that ${\cal G}_{d \uparrow} = {\cal G}_{d
  \downarrow}$.
If the upper arm of the ring is closed off (by adjusting $V_u$),
 the conductance $G_l$ through the lower
arm is proportional to\cite{meir92a}
 $ |t_L t_R|^2/(|t_L|^2 + |t_R|^2) \sum_\sigma {\rm Im} (t_{d\sigma})$.
Hence the amplitude of the oscillations in $G_{AB} $, normalized
to $G_l$, 
should be proportional to $|t_{d \sigma}| / {\rm Im } (t_{d\sigma}) 
= 1/\sin (\phi_{d\sigma})$, which
can be used as consistency check on assumptions (i) to (v).

Measurements of $\phi_{d\sigma}$ have so far been performed only for dots
without Kondo correlations, but even so the
$V_{g{}}$-dependence was interesting: $\phi_{d\sigma}$ increased by
$\pi$ whenever the dot was tuned through a Coulomb-blockade peak, as
expected for a Breit-Wigner type resonance.  It also
suffered a ``phase lapse'' (a drop by $-\pi$) between
consecutive peaks \cite{schuster}, which can be
explained by taking Coulomb interactions (at a mean-field level) on
the dot into account \cite{oreg97a,hackenbroich96a}. More
puzzling is the fact that the lapses persisted over many consecutive
valleys, but we do not wish to address this matter here
\cite{baltin99}.

{\em Kondo correlations.}--- Instead, we consider here a {\em
  single}\/ odd valley and study how $t_{d\sigma}$ is influenced by
Kondo correlations of the kind observed
recently\cite{goldhaber98a,cronenwett98a,goldhaber98b,schmid98a,simmel98}.
These were predicted \cite{glazman88a,meir93a} and interpreted
\cite{goldhaber98b} by using a standard model describing a localized
state (the $d$-level) coupled to a band of conduction electrons (both
the left and right sides of the base region; by assumption (v) we
henceforth neglect the influence of source and drain), namely the
much-studied and well-understood Anderson
model\cite{hewson-book,anderson,tsvelick83a}. The parameters of this
model, illustrated in Fig.~\ref{dot}(b), are: the energy
$\varepsilon_d$ of the $d$-level (measured relative to the chemical
potential of the base region, $\mu = 0$); the additional Coulomb
energy cost $U$ for having the $d$-level doubly occupied; the width of
the conduction band, which we take $\gg U$; and the
width\cite[(a)]{glazman88a} $\Gamma = \pi {\cal N}_{0}^{\rm tot}
(|t_L|^2 + |t_R|^2)$ of the $d$-level, where ${\cal N}_{0}^{\rm tot}$
is the combined density of states of all modes in the base region
which are coupled to the dot.

Sweeping the gate voltage $V_{g{}}$ into and through an ``odd valley''
in this model corresponds to sweeping the dot level $\varepsilon_d$
from above $\Gamma$ to below
$-(U+\Gamma)$, in the course of which the total average occupation of
the $d$-level, $\bar n_d$ ($ = 2 \bar n_{d\sigma}$), smoothly changes
from 0 to 2.  The valley center is at $\varepsilon_d = -U/2$, and its
two halves are related by particle-hole symmetry, with $\varepsilon_d
+ U/2 \to -( \varepsilon_d + U/2)$ implying $\bar n_d \to 2- \bar
n_d$.  As $\varepsilon_d$ is lowered through a half-valley towards
$-U/2$, three different regimes can be distinguished: (i) the
``empty-orbital'' regime $\varepsilon_d \gtrsim \Gamma$, in which
$\bar n_d \simeq 0$; (ii) the ``mixed-valence'' regime
$|\varepsilon_d| \lesssim \Gamma$, in which $\bar n_d$ begins to
increase due to strong charge fluctuations; (iii) the ``local-moment''
regime $-U/2 \le \varepsilon_d \lesssim -\Gamma$, in which $ \bar n_d$
approaches 1, so that the $d$-level acts like a localized spin. The
latter can give rise to Kondo correlations: as the temperature is
lowered below the Kondo temperature, a crossover scale given by $ T_K
= (U \Gamma/2 )^{1/2} e^{\pi \varepsilon_d (\varepsilon_d + U)/ 2
  \Gamma U} $ \cite{tsvelick83a}, the $d$-level density of states
$\rho_{d\sigma} (E)$ begins to develop a sharp peak near $E=0$ 
[dotted line in Fig.~\ref{dot}(b)], whose width is of order $T_K$
when $T \ll T_K$. This so-called Kondo resonance arises due to
coherent virtual transitions between the $d$-level and the conduction
band, which  ``screen'' the spin of the $d$-level
in such a way that the ground state is a spin singlet.  The resonance
strongly enhances the magnitude $|t_{d\sigma}|$ of the transmission
amplitude of electrons incident on the dot with energies $E \simeq 0$,
causing the dot's conductance in the local-moment regime of an odd
valley to be anomalously large at low $T$ and $V$, as seen in
\cite{goldhaber98a,cronenwett98a,goldhaber98b,schmid98a,simmel98}.
Typical dot parameters
\protect\cite{goldhaber98a,cronenwett98a,goldhaber98b} were $\Delta
\simeq 0.1-0.5$meV for the level spacing, and $\Delta/\Gamma
\simeq$1-3, $U/\Gamma \simeq $1-10, resulting in $T_K$'s between
45mK and 2K.

{\em Methods.---} To
study how Kondo correlations affect $\phi_{d\sigma}$,
we calculated $t_{d\sigma}$ via (\ref{transmission})
by three standard methods:

(a) For $T \gtrsim \Gamma (\gg T_K)$, where Kondo correlations are
weak, we used the {\em equations of motion}\/ (EOM) method; it
decouples higher into lower order Green's functions to yield an
analy\-tical expression for ${\cal G}_{d\sigma}(E)$ (Eq.\ (8) of
\onlinecite[(a)]{meir93a}, in which we calculated $\bar n_d$
self-consistently).

(b) For $T \! = \! V \! = \!0$, we have $t_{d\sigma} = - {\cal
  G}_{d\sigma}$; using well-known Fermi-liquid results for the
latter (Eqs.~(5.47) and (5.50) of Ref.~\onlinecite{hewson-book}), one
finds
\begin{equation}
  \label{eq:t_D-T=0}
  |t_{d\sigma}| = \sin (\bar n_{d \sigma} \pi ) \; , 
\qquad \phi_{d\sigma} = \bar n_{d \sigma} \pi .
\end{equation}
\noindent
\begin{figure}[htbp]
\begin{center}
    \epsfig{figure=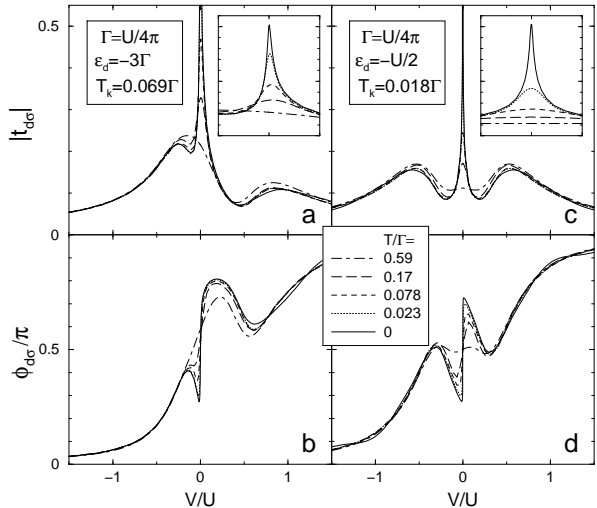,width=0.90\linewidth}
    \caption{Magnitude and phase of 
$t_{d\sigma}(V_{g{}}^{\rm fixed},V,T)$ (from NRG) 
as functions of $V/U$ and $T/\Gamma$, for $\Gamma \!=\! U/4 \pi $ and fixed
 $\varepsilon_d = -3 \Gamma$  (a,b) or 
 $ \varepsilon_d = -U/2$ (c,d).  
Insets show the range $|V| \le 20 T_K$; the $T\!=\!0$ peaks heights
are within 2{\%} of Eq.~(\ref{eq:t_D-T=0}).
\vspace*{-8mm} }
    \label{V}
\end{center}
\end{figure}
The second relation is the Friedel sum rule
\cite{friedel,langreth}. Thus, $t_{d\sigma}(V_{g{}})$ (for
$V=T=0$) is completely determined by ${\bar {n}_{d\sigma}}
(T=0)$, which for all $\Gamma, U, \varepsilon_d$ is known 
exactly from the {\em Bethe Ansatz}\/ (Eqs.~(8.2.47-48) of
\onlinecite{tsvelick83a}).

(c) For arbitrary temperatures $(\lesssim \Gamma)$, the only approach
which gives reliable results for ${\cal G}_{d\sigma} (E)$ for all
$\Gamma,U,\varepsilon_d$ is the {\em numerical renormalization
  group}\/ (NRG) \cite{costi94a,hewson-book}. It is designed to
calculate the density of states $\rho_{d\sigma} (E) \equiv -
\mbox{Im}\, {\cal G}_{d\sigma} (E) / \pi$, but this is sufficient to
determine ${\rm Re} \, {\cal G}_{d\sigma}(E)$ too, via a
Kramers-Kronig relation.

We calculated $t_{d \sigma}$ for two types of situations:

(1) {\em $V$-dependence.}--- Sweeping the source-drain voltage $V$,
with $V_{g{}}$ fixed in an odd valley (Fig.~\ref{V}), is the most
direct way of ``imaging'' the Kondo resonance, since by
Eq.~(\ref{transmission}) the $V$-dependence of the transmission
amplitude $t_{d\sigma}(V_{g{}}^{\rm fixed},V,T)$ reflects the
(thermally-smeared) $E$-dependence of ${\cal G}_{d\sigma}(E)$ (but
only if assumptions (i) and (v) hold \cite{coulomb,yacoby-priv}).  For
an asymmetric choice $\varepsilon_d = -3 \Gamma$ in the local-moment
regime [Fig.~\ref{V}(a,b)] and large temperatures, $|t_{d\sigma}|$
shows two broad peaks near $\varepsilon_d$ and $\varepsilon_d+U$, and
$\phi_{d\sigma}$ a weak phase lapse in between, as expected for two
not-very-well-separated single-particle resonances.  As $T$ is
lowered, a strong Kondo resonance in $|t_{d\sigma}|$ develops, as seen
in \cite{goldhaber98a,cronenwett98a,goldhaber98b,schmid98a,simmel98}.
Simultaneously, $\phi_{d\sigma}$ develops a novel sharp ``Kondo double
phase lapse'', because, intuitively speaking, it tends to lapse
between every two resonances, and now there are three, two broad and
one sharp.  These features become more pronounced the deeper
$\varepsilon_d$ lies in the local-moment regime, so much so that in
the symmetric case $\varepsilon_d = -U/2$ [Fig.~\ref{V}(c,d)] the
Kondo peak in $|t_{d\sigma}|$ and the double phase lapse in
$\phi_{d\sigma}$ are still faintly noticeable even for the highest
temperature shown ($T = 33 T_K$).  Encouragingly, these features might
thus be observable even at the center of an odd valley, where $T_K$ is
smallest
\begin{figure}[htbp]
\begin{center}
   \epsfig{figure=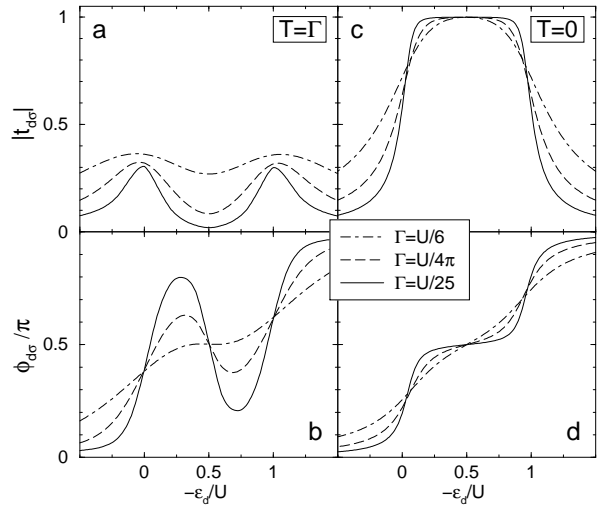,width=0.90\linewidth} \vspace*{-2mm}
    \caption{Magnitude and phase of $t_{d\sigma}(V_{g{}},0,T)$ as a
      function of $- \varepsilon_d/U$, for three values of $U/\Gamma$,
with (a,b) $T=\Gamma$ (from EOM); and (c,d) $T=0$ 
(from Bethe Ansatz).  \vspace*{-1cm}
}     \label{ed-U}
\end{center}
\end{figure}
\noindent
and  $T\lesssim T_K$ is hardest to
achieve.  If $\varepsilon_d$ is shifted from the local-moment into the
empty-orbital regime, the Kondo resonance merges with the lower broad
single-particle resonance and the double phase lapse for
$\phi_{d\sigma}$ disappears.

\begin{figure}[htbp]
\begin{center}
    \epsfig{figure=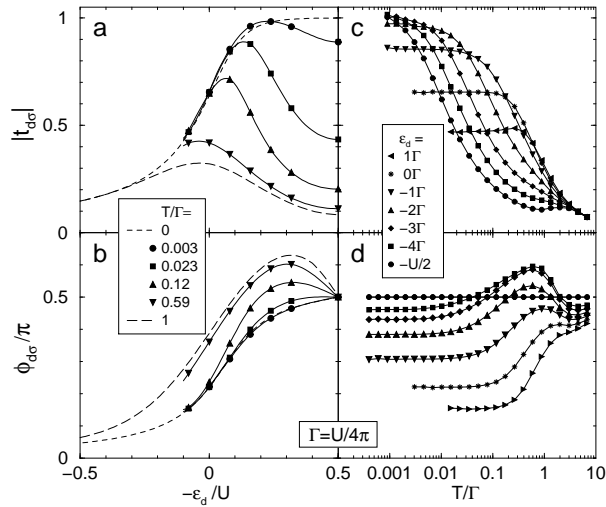,width=0.90\linewidth}
    \caption{Magnitude and phase of $t_{d\sigma}(V_{g{}},0,T)$,
for $\Gamma = U/4 \pi$: (a,b) as functions of 
$- \varepsilon_d/U$,  for 
$T=\Gamma$ (long-dashed, from EOM),
$T=0$ (short-dashed, from Bethe Ansatz), and four intermediate
values of  $T/\Gamma$; 
(c,d) as functions of $T/\Gamma$, for various $\varepsilon_d$. 
The symbols represent points calculated using the NRG;
they  are  connected by spline fits. 
\vspace*{-6.5mm}}
    \label{ed-T}
\end{center}
\end{figure}

(2) {\em $V_g$-dependence.}--- 
Sweeping $V_{g{}}$ through an odd valley, with $V=0$
(Figs.~\ref{ed-U} and \ref{ed-T})  probes  the
low-energy window $|E| \lesssim T$ within which the Kondo resonance
shoots up for $T \lesssim T_K$;  the magnitude of $t_{d\sigma}
(V_{g{}},V=0,T)$ thus reflects the weight of the Kondo resonance.  For
large $T$ (say $\simeq \Gamma$, i.e.\ negligible Kondo correlations)
and a small $ \Gamma/U$, $t_{d\sigma}$ shows the familiar behavior
[Fig.~\ref{ed-U}(a,b)] experimentally observed in
Ref.~\onlinecite{schuster}: its magnitude $|t_{d\sigma}|$ has
well-resolved Coulomb-blockade peaks near 0 and $-U$, at each of which
its phase $\phi_{d\sigma}$ rises (by almost $\pi$), with a
significant phase lapse in the valley in between.  The larger
$\Gamma/U$, the less sharp these features, since the peaks
increasingly overlap. As $T$ is lowered, the Kondo
resonance develops throughout the local-moment regime, leading to a
dramatically different picture at $T=0$ [Fig.~\ref{ed-U}(c,d)]:
$|t_{d\sigma}|$ has just one, much higher peak, and $\phi_{d\sigma}$
increases {\em monotonically}\/ from 0 to $\pi$; by
Eq.~(\ref{eq:t_D-T=0}), both reflect the monotonic change in the
occupation $\bar n_d$ of the $d$-level as $\varepsilon_d$ is swept.
If $\Gamma/U$ decreases, the extent (in units of $\Gamma$) of the
$\bar n_d \simeq 1$ local-moment regime increases, so that both
$|t_{d\sigma}|$ and $\phi_{d\sigma}$ develop flat plateaus around
$\varepsilon_d = -U/2$.

The $\phi_{d\sigma} = \pi/2$ plateau is the manifestation of the
famous $\pi/2$ Kondo phase shift mentioned earlier. It arises because
the local spin is screened into a singlet at $T=0$. Since {\em no}\/
spin-flip scattering occurs, a Fermi-liquid description of the system
is possible\cite{nozieres}: to low-energy Fermi-liquid quasiparticles
scattering off the singlet Kondo resonance, it looks like a {\em
  static}\/ (as opposed to dynamical) impurity, which scatters them
without
 randomizing their phase, and which is strongly
repulsive, causing a resonance phase shift\cite{friedel,langreth} of
$\pi/2$.
\noindent

{\em $T$-dependence.}---
Figs.\ \ref{ed-T}(a,b) show the crossover from
Figs.\ \ref{ed-U}(a,b) to \ref{ed-U}(c,d) as the temperature is
lowered from $\Gamma$ to 0, for $\varepsilon_d \ge
-U/2$ and $\Gamma=U/ 4 \pi$ (this value was also used in
\protect\onlinecite{costi94a}, whose Figs.~5 to 10 show how the
Kondo peak changes correspondingly). Figs.~\ref{ed-T}(c,d) show the
same crossover, but now with $T/\Gamma$ on the horizontal axis.  As $T$ 
approaches $\Gamma$ from above, $\phi_{d\sigma}$ initially rises
if $\varepsilon_d \lesssim 0$ and drops if $\varepsilon_d > 0$, because the
phase rise in Fig.~\ref{ed-U}(b) sharpens.  As $T$ is decreased
further, $\phi_{d \sigma}$ decreases for {\em all}\/ $\varepsilon_d$,
reflecting the Kondo suppression, shown in Fig.~\ref{ed-T}(b), of the
phase lapse.  Thus, a maximum (instead of a low-$T$ saturation) in
$\phi_{d\sigma}(T)$ for $-U/2 < \varepsilon_d \lesssim 0$ would 
signify the onset of Kondo correlations.

{\em $H$-dependence}.--- A strong magnetic field that lifts the
spin degeneracy of the local level by $\varepsilon_\uparrow -
\varepsilon_\downarrow = \Delta_h$ will split the Kondo resonance into
two sub-resonances, separated by 
$\simeq \Delta_h$  \cite{meir93a,cronenwett98a}, 
while strongly reducing
their combined weight, and thereby also the spectral weight at $E
\simeq 0$.  Thus, with increasing $\Delta_h$, type (2)
measurements should behave similarly as for increasing $T$, which also
reduces the spectral weight at $E \simeq 0$; and type (1) measurements
should show two Kondo peaks (of reduced height) in $|t_{d\sigma}|$ and
a Kondo triple phase lapse in $\phi_{d\sigma}$ \cite{vD}.

To summarize, we studied phase-coherent transport of electrons traversing a
strongly-interacting environment (the dot-lead system) that is tunable from
being weakly correlated at high temperatures through a strongly-correlated
crossover regime to a Fermi liquid\cite{nozieres} at sufficiently low
temperatures. We identified three ``smoking guns'' for Kondo correlations in
the behavior of $\phi_{d\sigma}$: the Kondo double phase lapse in
Figs.~\ref{V}(b,d); the $\pi/2$ plateau in Fig.~\ref{ed-U}(d); and the maxima
in Fig.~\ref{ed-T}(d).  The experimental observation of the $\pi/2$ plateau
would constitute the first direct measurement of the $\pi/2$
Kondo phase shift 
predicted more than 30 years ago.

We thank R. Blick, R. Fazio, Y. Gefen, D. Goldhaber-Gordon, J.
G\"ores, L. Glazman, A. Finkel'stein, M. Heiblum, J. Kroha, K.
Matveev, T. Pruschke, H. Schoel\-ler, G. Sch\"on, A. Yacoby, J. Yang
and A. Zawadowski for discussions.  J.v.D. was supported by SFB195 of
the DFG and the German-Israeli-Project of the BMBF; U.G. by the DFG
Graduiertenkolleg; and Y.O. by the NSF DMR grants 94-16910, 96-30064,
97-14725 and 98-09363.

${}^\ast$
Present address:
Institut Laue-Langevin, BP156, F-38042 Grenoble Cedex 9, France\vspace*{-7.5mm}.

\widetext

\end{document}